# COMMISSIONING OF DEDICATED FURNACE FOR Nb$_3$Sn COATINGS OF 2.6 GHZ SINGLE CELL CAVITIES

P.A. Kulyavtsev[1,2], G. Eremeev[*1], S. Posen[1], B. Tennis[1], J. Zasadzinski[2]
[1]Fermi National Accelerator Laboratory, Illinois, USA
[2]Illinois Institute of Technology, Chicago, USA.

## ABSTRACT

We present the results of commissioning a dedicated furnace for Nb$_3$Sn coatings of 2.6 GHz single cell cavities. Nb$_3$Sn is a desired coating due to its high critical temperature and smaller surface resistance compared to bulk Nb. Usage of Nb$_3$Sn coated cavities will greatly reduce operating costs due to decreased dependance on cryo cooling. Tin is deposited by use of a tin chloride nucleation agent and tin vapor diffusion. Analysis of the resultant coating was performed using SEM/EDS to verify successful formation of Nb$_3$Sn. Witness samples in line of sight of the source were used in order to understand the coating efficacy.

## INTRODUCTION

Nb$_3$Sn-coated cavities achieve accelerating gradients E$_{acc}$ above 10 MV/m and quality factors in excess of $10^{10}$ at 4 K [1–5]. High quality factors of Nb$_3$Sn-coated cavities at 4 K are the enabling technology for compact cryomodules for industrial accelerators [6]. Several projects look to exploit this technology for various industrial applications [7,8]. Given that the superheating critical field and superconducting transition temperature of Nb$_3$Sn is higher than that of niobium, Nb$_3$Sn superconducting material has the potential to sustain accelerating gradients twice that of niobium cavities and quality factors close to $5 \cdot 10^{10}$, which will significantly reduce the operating and capital cost of future accelerators. Research and development efforts are ongoing to understand and improve vapor diffusion deposition techniques to realize they potential of Nb$_3$Sn material.

Several institutions are pursuing Nb$_3$Sn coatings and research is ongoing to understand material limitations and to optimize coating process [9–12]. Nb$_3$Sn coatings at Fermilab are done in the large coating system designed to coat multicell cavities. The coating system is constantly used to coat cavities for various projects. In order to enable Nb$_3$Sn coating research on small cavities and samples in parallel with multicell cavity coating, we refurbish and commission for Nb$_3$Sn coating an old furnace. Unlike the larger Nb$_3$Sn system, which is routinely used for coating of Nb$_3$Sn films on single and multi-cell cavities [4], the new system will be used to coat smaller single-cell cavities and for samples studies. This contribution presents the results from the commissioning of the new coating system.

* grigory@fnal.gov

## EXPERIMENTAL SETUP

For these Nb$_3$Sn studies, the old IVI Model 3312-1212-120V furnace was adopted. This is the horizontal front loading furnace that features 12" x 12" x 15" hot zone. The furnace heaters as well as the inner shields are made out of molybdenum. The furnace is controlled by PLC connected to a computer and can reach up to 1200 $°C$. The furnace is evacuated with an oil-free mechanical pump and a cryopump and can reach down to $10^{-8}$ Torr range cold. The temperature is monitored with four molybdenum-sheathed type C thermocouples.

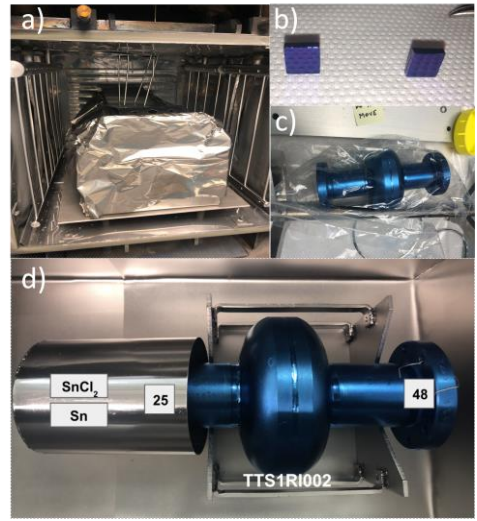

Figure 1: *Experimental setup for Nb$_3$Sn coating. [a] Assembled experimental setup in the hot zone of the furnace. Note niobium foil covering the assembly and thermocouples at the top. [b,c] Niobium samples (b) and niobium cavity (c) prepared for the coating. Note that the blue color is due to 30 V anodizing. [d] Niobium cavity assembled for the coating inside niobium box on the niobium support fixture.*

For Nb$_3$Sn coating, a cavity is placed inside a niobium box onto niobium supports, fig. 1. Two samples, one on each end of the cavity, are mounted with niobium wires. One end of the cavity is then assembled with a molybdenum cup, which holds Sn and SnCl$_2$. The other end is left open. For the first coating of 2.6 GHz cavity, 2 g of Sn and 0.5 g of SnCl$_2$ were used. The assembly is then covered with niobium foil. As part of the preparation for the first Nb$_3$Sn coating on SRF cavity, both the cavity and the samples were anodized for 30 V. The coating was done at 1150 °C for 3 hours preceded by the nucleation step at 500 °C. 1.6 g of Sn was consumed in the coating.

## MEASUREMENTS AND RESULTS

After the coating, the cavity surface and the samples were visually inspected, Fig.2. The inner surface of the cavity appeared to be well coated with a thick Nb$_3$Sn layer. The sample that was placed far from the Sn and SnCl$_2$ source appeared uniformly coated. On the sample that was close to the source, some variation in the coating could be visually observed.

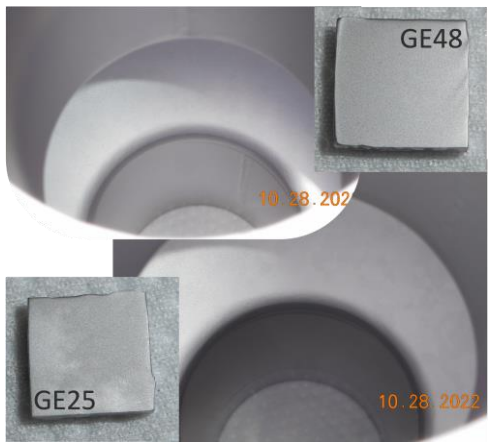

Figure 2: *Images of the inner surface of the coated cavity and samples. Coated cavity half cells are shown in the top left and bottom right corners of the cavity. In the top right corner the sample GE048, mounted far from the source, is shown. In the bottom left corner the sample GE025, mounted close to the source, is shown.*

### Sample analysis

Samples were studied with FEI Helios dual beam system and Oxford EDS. Nearside (GE025) and far side (GE048) samples were compared to understand the quality of the coating, Fig.3. EDS analysis established that atomic percentage of Nb to Sn was in the desired range. Visual comparison of the samples revealed more frequent pits or holes on the surface of the far side sample (GE048). Relative increase in these features was previously linked with Sn deficiency in the coating process [13], which is supported by this data. The dual beam system was used to perform FIB on the sample. FIB crosssections show that the coating was 2-3 $\mu$m thick consistent with the visual inspection, Fig.3.

### Cavity testing

Nb$_3$Sn-coated 2.6 GHz cavity was cleaned with HPR and assembled in cleanroom for cryogenic testing in vertical dewar. For the cryogenic test the cavity was instrumented with fluxgates, Cernox temperature sensors, and the magnetic field compensation coil, Fig. 5. The test was conducted with the cavity under static vacuum. Because of bi-metallic nature of Nb$_3$Sn coating on niobium, the dewar was slowly cooled down with the cavity in the compensated field. The field reading of the fluxgate, which was used for field compensation, read about 0 mG. The other fluxgate reading was close to 8 mG, when the cavity temperature crossed about 18 K. The temperature difference across four temperature sensors was about 100 mK, when the average temperature of the cavity crossed 18 K during cooldown. In Fig. 6, cavity test results are shown. At 4.4 K, the cavity exhibit the low field quality factor of about $5 \cdot 10^{10}$. Quality factor degraded to about $5 \cdot 10^{10}$ at E$_{acc}$ = 11.5 quench field. At 2.0 K, the cavity exhibit the low field quality factor of above $2 \cdot 10^{10}$. Quality factor degraded to about $8 \cdot 10^9$ at E$_{acc}$ = 11.8 quench field.

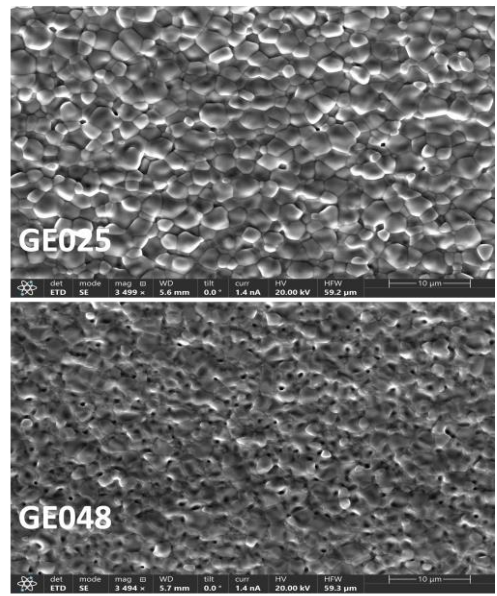

Figure 3: *SEM inspection of Nb$_3$Sn-coated samples. [top] sample mounted close to Sn and SnCl$_2$ source. [bottom] sample mounted far from the source. Note more pit-like features in the sample that was far from the source.*

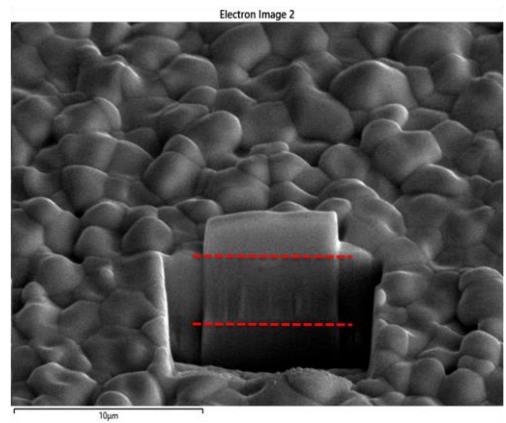

Figure 4: *FIB crosssection of Nb$_3$-coated sample. Red dashed lines indicate the thickness of Nb$_3$Sn coating.*

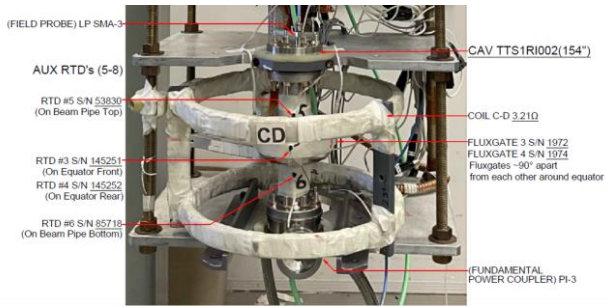

Figure 5: *Cavity prepared for vertical testing on the vertical test stand. Note the magnetic field compensation coil, cernox temperature sensors, and the fluxgates.*

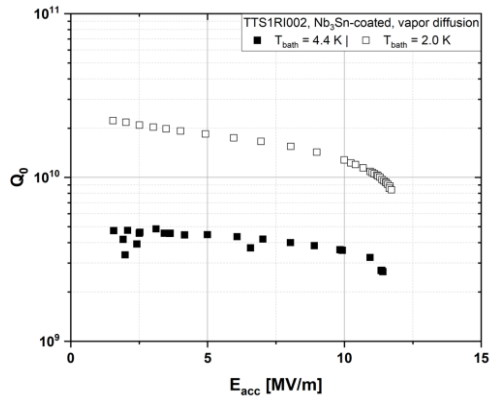

Figure 6: *Cavity test results. [filled black squares] quality factor vs field at 4 K. [empty black squares] quality factor vs field at 2 K.*

## CONCLUSION

In this contribution we presented the first results of $Nb_3Sn$ coating on a 2.6 GHz cavity in the recently refurbished furnace. The coating appearance on the cavity was uniform. However, the detailed analysis of samples coated with the cavity presented evidence of low Sn flux during the coating. In the cryogenic RF test, the cavity showed $Q_0$ of about $5 \cdot 10^9$ and was limited to about 11.5 MV/m by quench at 4 K. At 2 K, the cavity $Q_0$ was about $2 \cdot 10^{10}$ at low fields and the cavity was limited to about 12 MV/m at 11.8 K. With these promising results, we are continuing $Nb_3Sn$ coating optimization using a mock cavity with the goal to control and improve Sn flow.

## ACKNOWLEDGMENTS

This manuscript has been authored by Fermi Research Alliance, LLC under Contract No. DE-AC02-07CH11359 with the U.S. Department of Energy, Office of Science, Office of High Energy Physics. This material is based upon work supported by the U.S. Department of Energy, Office of Science, Office of Nuclear Physics.

## REFERENCES

[1] S. Posen and M. Liepe, Proc. 16th Int. Conf. RF Supercond., (2013), pp. 666-669

[2] G.V. Eremeev et al., in Proc. 17th Int. Conf. RF Super., (2015), pp. 505-511

[3] S. Posen, M. Liepe, and D. L. Hall, Appl. Phys. Lett. 106, 082601 (2015)

[4] S. Posen and D.L. Hall, Supercond. Sci. Technol. 30 033004 (2017)

[5] S. Posen, TTC meeting, Vancouver, Canada (2019)

[6] R. Kephart et al. Proc. 17th Int. Conf. RF Supercond, (2015)

[7] G. Ciovati et al., Phys. Rev. AB 26, 044701 (2023)

[8] R.C. Dhuley et al, Phys. Rev. Accel. Beams 25, 041601 (2022)

[9] C. Dong et al., Phys. C, Volume 600, (2022) 1354107

[10] K. Takahashi et al., Proc. 20th Int. Conf. RF Supercond., (2021), pp. 23 - 26

[11] U. Pudasaini et al., Proc. 20th Int. Conf. RF Supercond., (2021), pp. 516 - 521

[12] R. D. Porter et al., Proc. 20th Int. Conf. RF Supercond., (2021), pp. 6 - 10

[13] U. Pudasaini et al., 2020 Supercond. Sci. Technol. 33 045012